\documentclass[floatfix, aip, jap, superscriptaddress, reprint]{revtex4-1}

\usepackage{graphicx}               
\usepackage{bm}                     
\usepackage{amsfonts,amssymb}       
\usepackage{dcolumn}                
\usepackage{rotating}
\usepackage{multirow}

\newcommand*\pct{\scalebox{.9}{\%}}

\begin{document}

\title{
Activated lone-pair electrons lead to low lattice thermal conductivity:\\
a case study of boron arsenide
}

\author{Guangzhao~Qin}
\email{qin.phys@gmail.com}
\affiliation{Institute of Mineral Engineering, Division of Materials Science and Engineering, Faculty of Georesources and Materials Engineering, RWTH Aachen University, Aachen 52064, Germany}
\affiliation{Department of Mechanical Engineering, University of South Carolina, Columbia, SC 29208, USA}
\author{Zhenzhen~Qin}
\affiliation{International Laboratory for Quantum Functional Materials of Henan, and School of Physics and Engineering, Zhengzhou University, Zhengzhou 450001, China}
\author{Huimin~Wang}
\affiliation{College of Engineering and Applied Science, Nanjing University, Nanjing 210023, China}
\affiliation{Department of Mechanical Engineering, University of South Carolina, Columbia, SC 29208, USA}
\author{Ming~Hu}
\email{hu@sc.edu}
\affiliation{Department of Mechanical Engineering, University of South Carolina, Columbia, SC 29208, USA}

\date{\today}


\begin{abstract}
Reducing thermal conductivity ($\kappa$) is an efficient way to boost the
thermoelectric performance to achieve direct solid-state conversion to
electrical power from thermal energy, which has lots of valuable applications
in reusing waste resources.
In this paper, we propose an effective approach for realizing low $\kappa$ by
introducing lone-pair electrons or making the lone-pair electrons
stereochemically active through bond nanodesigning.
As a case study, by cutting at the (111) cross section of the three-dimensional
(3D) cubic boron arsenide ($c$-BAs), the $\kappa$ is lowered by more than one
order of magnitude in the resultant two-dimensional (2D) system of graphene-like
BAs ($g$-BAs) due to the stereochemically actived lone-pair electrons.
However, this does not naturally happen to all materials.
For instance, breaking the perfect octahedral coordination of 3D diamond as in
the 2D graphene adversely enhances thermal transport.
The underlying mechanism is analyzed based on the comparative study on the
thermal transport properties of $g$-BAs, $c$-BAs, graphene, and diamond
($c$-BAs $\to$ $g$-BAs \emph{vs.}\ diamond $\to$ graphene).
Furthermore, deep insight into the electronic origin is gained by performing
fundamental analysis on the electronic structures.
Similar concept can be also extended to other systems with lone-pair electrons
beyond BAs, such as group III-V compounds (\emph{e.g.}\ BN, AlN, GaN,
\emph{etc}), where a strong correlation between $\kappa$ modulation and
electronegativity difference for binary compounds is found.
Thus, the lone-pair electrons combined with a small electronegativity difference
could be the indicator of lowering $\kappa$ through bond nanodesigning to change
the coordination environment.
The proposed approach for realizing low $\kappa$ and the underlying mechanism
uncovered in this study would largely benefit the design of thermoelectric
devices with improved performance, especially in future research involving novel
materials for energy applications.
\end{abstract}

\maketitle


\section{Introduction}
%
%

Due to the ability of firsthand solid-state conversion to electrical power from
thermal energy, especially for waste heat reusing, thermoelectrics have
attracted a lot of attention in recent years\cite{Nature.2012.489..414}.
Thermoelectrics have lots of valued applications in recovering resources and
thus may make crucial contributions to the crisis of environment by solving
energy problems\cite{QINsrep046946}.
Moreover, thermoelectrics possess the advantages of having no moving components
and being environmentally friendly compared to traditional mechanical heat
engines.
Generally, the thermoelectric efficiency and performance can be characterized by
a dimensionless figure of merit\cite{Science.2016.351.6269.141-144}
\begin{equation}
\label{eq:ZT}
ZT = S^2\sigma T/\kappa\ ,
\end{equation}
where $S$, $\sigma$, $T$ and $\kappa$ are thermopower (Seebeck coefficient),
electrical conductivity, absolute temperature and total thermal conductivity,
respectively.
The commercial applications in industry of thermoelectric devices are currently
limited by the low ZT merit.
To approach the Carnot coefficient as closely as possible, a high energy
generation efficiency is necessary, which corresponds to a large ZT merit.
Based on the definition [Eq.~(\ref{eq:ZT})], lowering the $\kappa$ would be more
efficient to boost the ZT merit due to the inversely proportional
relation\cite{PhysRevB.89.054310}.

Previous theoretical studies predicted that cubic boron arsenide ($c$-BAs) in
the bulk form has an exceptionally high $\kappa$ over 2000\,W/mK, which is
comparable to the bulk carbon crystals (diamond) with record highest $\kappa$
\cite{PhysRevLett.111.025901}.
The ultra-high $\kappa$ of $c$-BAs was analyzed to be resulted from the large
phonon band gap between acoustic and optical phonon branches together with the
bunching of the acoustic phonon branches, which reduce phonon-phonon scattering
\cite{PhysRevLett.111.025901}.
The features of the phonon dispersion of $c$-BAs analyzed based on
first-principles calculations are then confirmed by experimental measurements
based on inelastic $x$-ray scattering\cite{Phys.Rev.B.2016.94.22.220303}.
By considering the phonon-phonon scattering involving four phonons, Feng
\emph{et al.}\cite{PhysRevB.2017.96.161201}
found that the $\kappa$ of $c$-BAs reduces from 2200 to 1400\,W/mK, which was
recently confirmed by experimental studies
\cite{Science.2018.Yongjie.Hu.5522, Science.2018.cahill.8982, Science.2018.zhifeng.ren.7932}.
Experimental measurements reveal that the $\kappa$ of BAs can be suppressed by
the arsenic deficiency or vacancy in the BAs
sample\cite{Appl.Phys.Lett..2015.106.7.074105}
and the phonon-boundary scattering in BAs microstructures
\cite{PhysRevB.88.214303, Appl.Phys.Lett..2016.108.20.201905}.
However, the obtained $\kappa$ of BAs is still too high that limits its
potential applications in thermoelectrics, despite that the Seebeck coefficient
and thermoelectric power factor of BAs is comparable to those of bismuth
telluride, \cite{Appl.Phys.Lett..2016.108.20.201905}
which is one of the most commonly used thermoelectric materials.
Thus, it would be meaningful if one can find an effective approach to lower the
intrinsic $\kappa$ of BAs to benefit its applications in thermoelectrics.
Moreover, the approach that makes such a high $\kappa$ material applicable for
thermoelectrics would also largely benefit the design of thermoelectric devices
with improved performance by lowering $\kappa$, especially in the future
research involving novel materials for energy applications.

\begin{figure}[tb]
    \centering
    \includegraphics[width=1.00\linewidth]{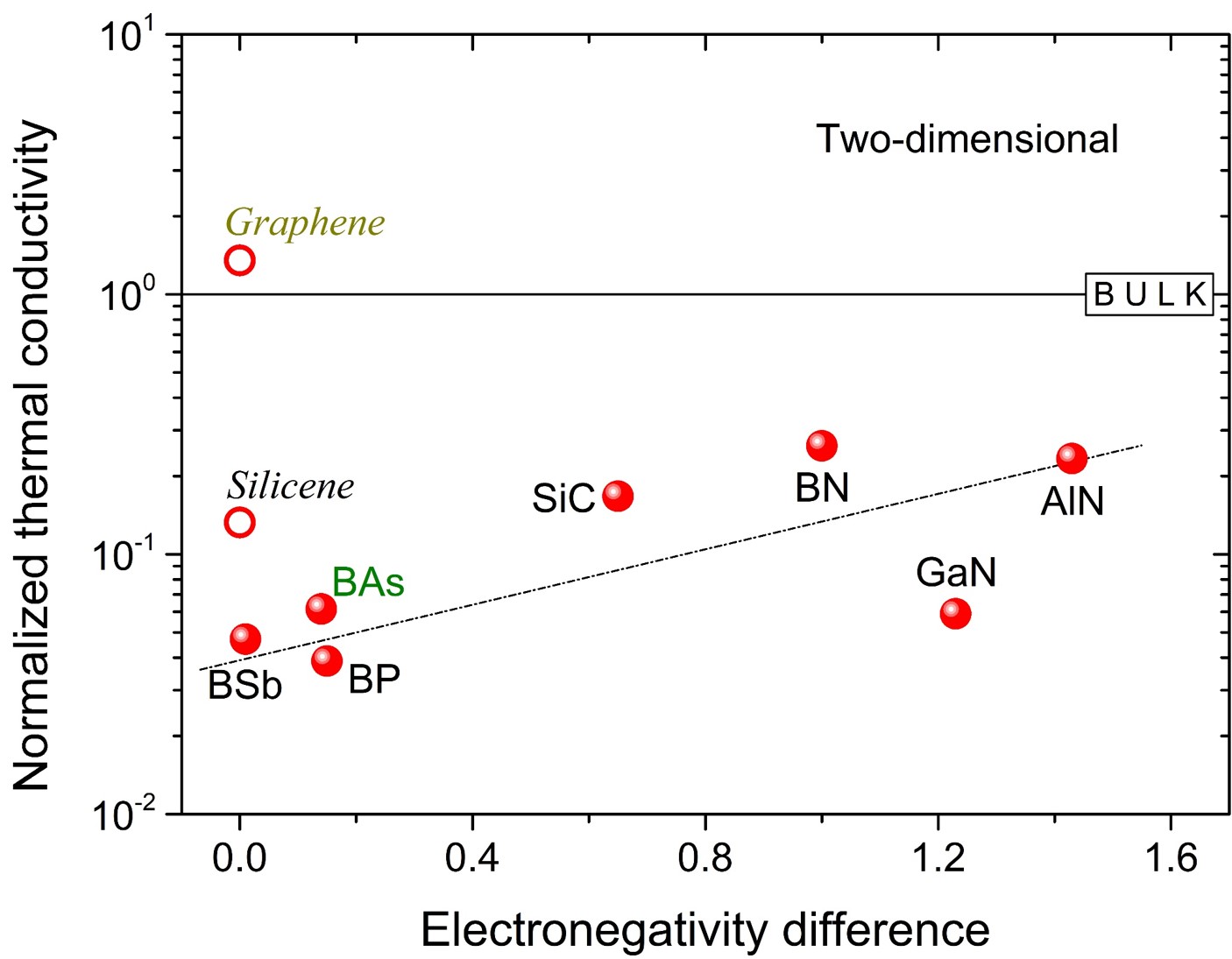}
\caption{\label{fig:compare}
The $\kappa$ of typical two-dimensional materials (graphene, silicene, BN, AlN,
GaN, BP, BAs, BSb, and SiC) at 300\,K, which are normalized to the bulk
counterparts, respectively.
The two-dimensional materials possess lower $\kappa$ compared to the bulk
counterparts, except graphene.
The dot line is for an eye guide.
The specific data can be found in Supplemental Table 1.
}
\end{figure}

\begin{figure}[tb]
    \centering
    \includegraphics[width=1.00\linewidth]{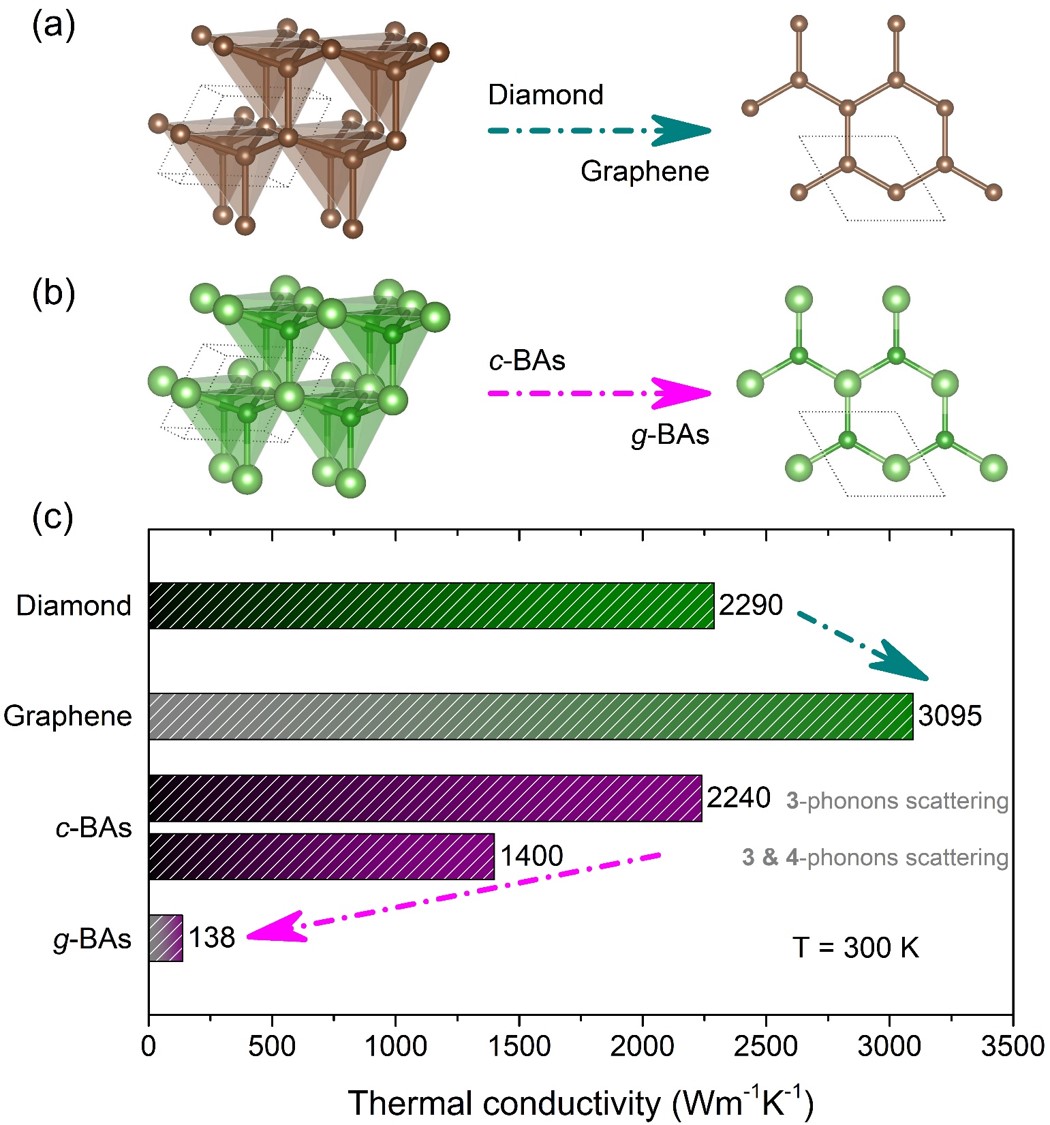}
\caption{\label{fig:kappa}
The similarity of the transformation from 3D cubic to 2D honeycomb planar
geometry structures ($c$-BAs $\to$ $g$-BAs \emph{vs.}\ diamond $\to$ graphene)
is in contrast to the opposite $\kappa$ variation.
When transforming from 3D into 2D, the $\kappa$ of BAs is found to be
anomalously lowered by more than one order of magnitude.
(a) The structure of graphene in 2D is the (111) cross section of the structure
of diamond in 3D, which is planar due to the $sp^2$ hybridization of carbon
atoms.
(b) The $g$-BAs to $c$-BAs is like graphene to diamond.
(c) The comparison of $\kappa$ of diamond, graphene,
$c$-BAs,\cite{PhysRevLett.111.025901, PhysRevB.2017.96.161201} and $g$-BAs.
}
\end{figure}

In this paper, we propose an effective approach for realizing low $\kappa$ by
bond nanodesigning to make the lone-pair electrons stereochemically active.
As a result, much lower $\kappa$ can be generally achieved, except the case of
graphene (Fig.~\ref{fig:compare}).
As a specific case study, when transforming the three-dimensional (3D) $c$-BAs
into the two-dimensional (2D) graphene-like BAs ($g$-BAs), the $\kappa$ is found
to be lowered by more than one order of magnitude (Fig.~\ref{fig:kappa}) due to
the stereochemically actived lone-pair electrons.
The underlying mechanism is analyzed based on the comparative study on the
thermal transport properties of $g$-BAs, $c$-BAs, graphene, and diamond,
considering the similarity of the transformation from 3D cubic to 2D honeycomb
planar geometry structures ($c$-BAs $\to$ $g$-BAs \emph{vs.}\ diamond $\to$
graphene) but the opposite trend for the $\kappa$ modulation.
Moreover, deep insight into the electronic origin is gained by performing
fundamental analysis on the electronic structures.


\begin{figure*}[tb]
    \centering
    \includegraphics[width=1.00\linewidth]{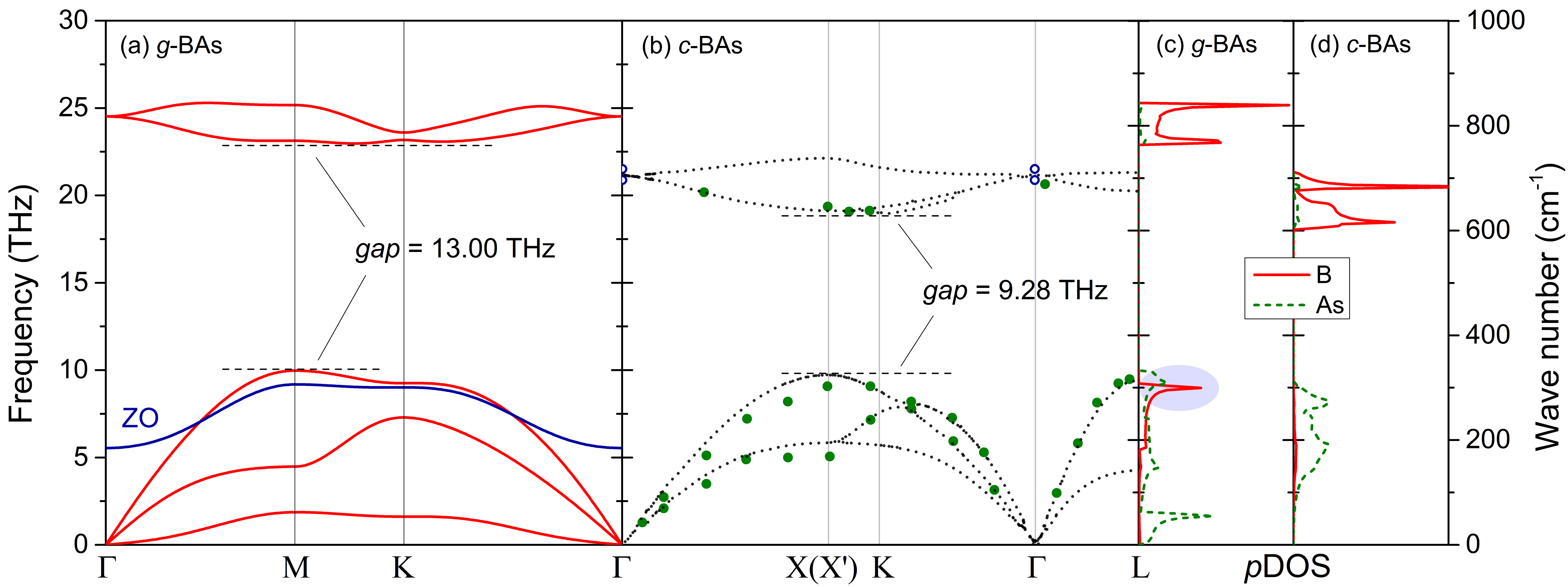}
\caption{\label{fig:dispersion}
The comparison of phonon dispersion of boron arsenide in 2D ($g$-BAs) with 3D
($c$-BAs).
(a) The phonon dispersion of $g$-BAs along the high-symmetry points, where the
\emph{out-pf-plane} flexural acoustic (FA) phonon branch shows a good quadratic
behavior (Supplemental Figure 1).
There exists a huge gap of 13.00\,THz between the optical and acoustic phonon
branches.
The $z$-direction optical (ZO) phonon branch, which is below the gap, is
highlighted in blue.
(b) The phonon dispersion of $c$-BAs along the high-symmetry points, where dot
lines in black are from theoretical calculations\cite{PhysRevLett.111.025901}
and blue/green points are from experimental
measurements\cite{Phys.Rev.B.2016.94.22.220303,
Phys.Rev.Lett..1994.73.18.2476-2479}.
The gap between the optical and acoustic phonon branches is 9.28\,THz.
(c,d) The partial density of states ($p$DOS) of (c) $g$-BAs and (d) $c$-BAs,
where the contribution from boron (B) atoms to ZO is highlighted with a colored
ellipse.
}
\end{figure*}

\begin{figure}[!b]
    \centering
    \includegraphics[width=1.00\linewidth]{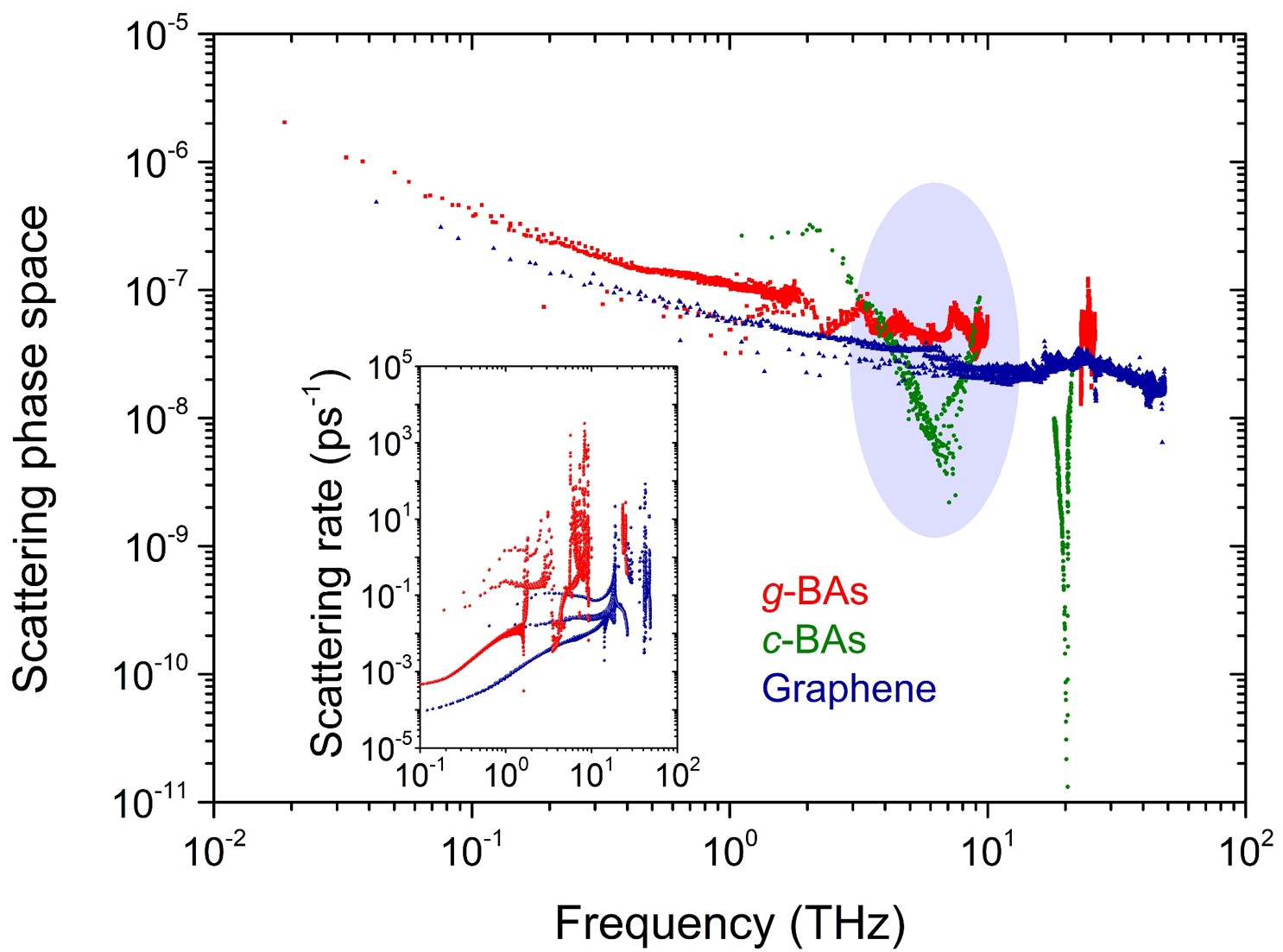}
\caption{\label{fig:P3}
Comparison of the phonon-phonon scattering phase space of $g$-BAs with $c$-BAs
and graphene.
The colored ellipse highlights the larger scattering phase space of $g$-BAs than
$c$-BAs in the frequency range of 5-10\,THz, where the ZO phonon branch lies.
(Inset) The comparison of scattering rate between $g$-BAs and graphene.
}
\end{figure}

\begin{figure*}[tb]
    \centering
    \includegraphics[width=0.95\linewidth]{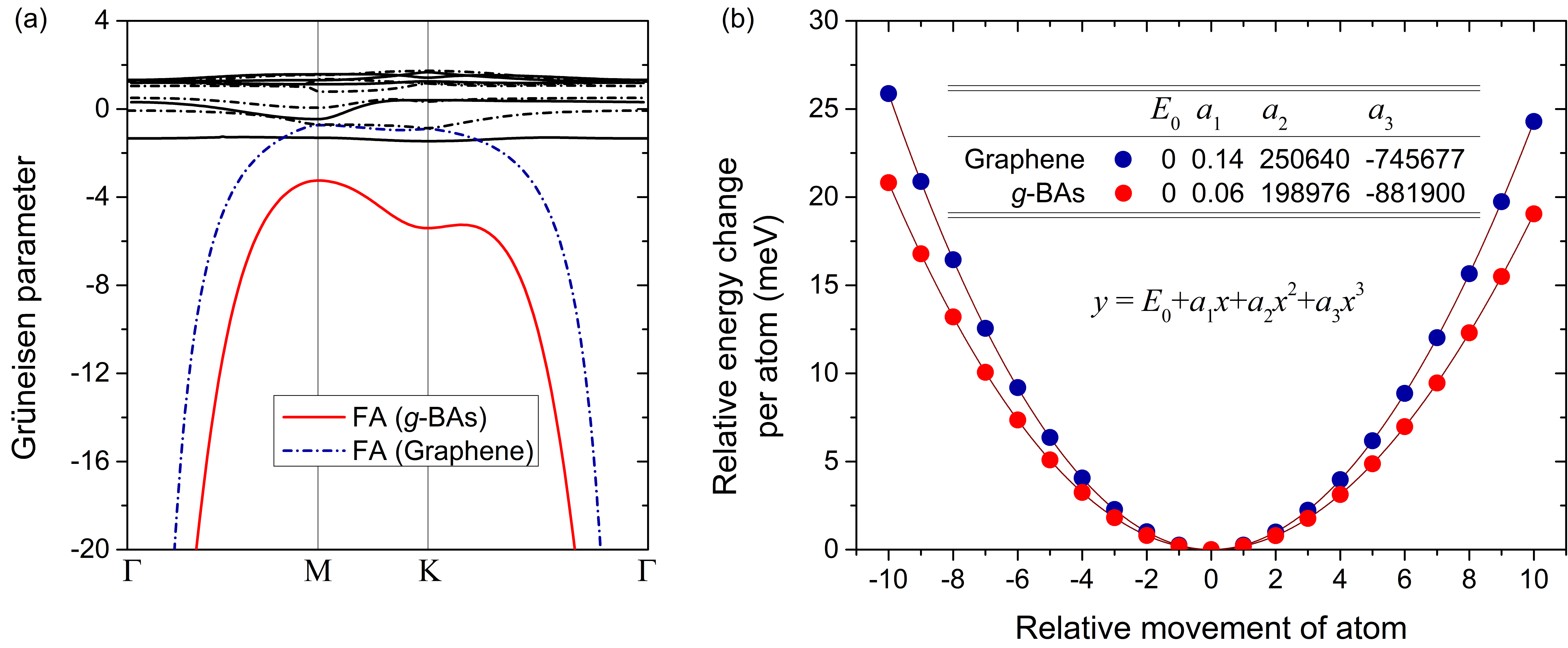}
\caption{\label{fig:anharmonicity}
Strong phonon anharmonicity in $g$-BAs.
(a) Comparison of Gr\"uneisen parameters between $g$-BAs and graphene.
The colored lines highlight the Gr\"uneisen parameters of FA phonon branch.
(b) Comparison of potential energy wells between $g$-BAs and graphene.
The atom is moved along the bonding direction.
Points are from first-principles calculations and lines are fittings to the
formula shown on site.
Inset table: The fitted parameters for $g$-BAs and graphene, respectively.
}
\end{figure*}

\begin{figure*}[tb]
    \centering
    \includegraphics[width=0.90\linewidth]{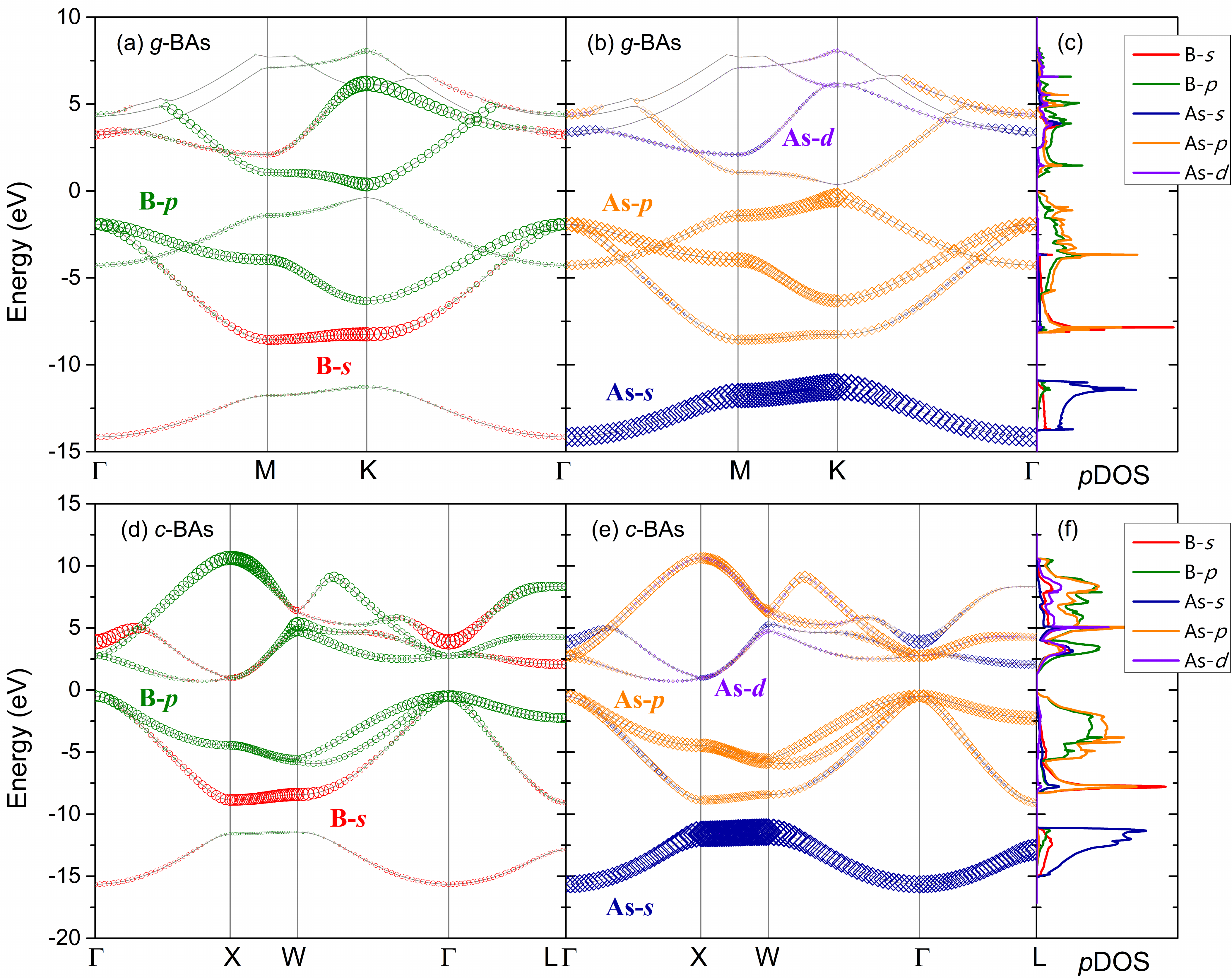}
\caption{\label{fig:band}
Orbitals projected electronic structures, revealing the non-bonding lone-pair
As-$s$ electrons.
(a,b,c) The orbitals projected (a,b) electronic band structures and (c) density
of states (DOS) for $g$-BAs.
The electronic structures are projected to (a) B-$s/p$ and (b) As-$s/p/d$
orbitals.
(d,e,f) Similar figures for $c$-BAs in comparison.
}
\end{figure*}

\section{Results and discussions}

\subsection{The anomalously low $\kappa$ of $g$-BAs}
%
By cutting the 3D cubic structure of $c$-BAs at the (111) cross section, $g$-BAs
can be obtained with similar planar honeycomb structure as graphene ($c$-BAs
$\to$ $g$-BAs \emph{vs.}\ diamond $\to$ graphene).
The $\kappa$ of $g$-BAs is obtained to be 137.70\,W/mK, which is more than one
order of magnitude lower than that of graphene (3094.98\,W/mK).
Note that only 3-phonons scattering is considered here for simplicity.
The $\kappa$ of $g$-BAs could be further lowered if 4-phonons scattering is
included.
The \emph{in-plane} longitudinal acoustic (LA), transverse acoustic (TA) and
\emph{out-pf-plane} flexural acoustic (FA) phonon branches contribute 28.5\pct,
43.1\pct\ and 26.9\pct, respectively.
The $\kappa$ of $g$-BAs are 89.3 and 137.7 W/mK before
and after iteration, respectively.
The large difference in the RTA and iteration result means that the proportion
of N-process could be large and there exists strong phonon hydrodynamics in
$g$-BAs.

The lower $\kappa$ of $g$-BAs than graphene is very intriguing considering the
similarity of their planar honeycomb geometry structures
[Fig.~\ref{fig:kappa}(a,b)].
In particular, the $\kappa$ of $c$-BAs is comparable to diamond
[Fig.~\ref{fig:kappa}(c)]\cite{PhysRevLett.111.025901}, both of which share the
same cubic structures and are the 3D counterparts of $g$-BAs and graphene,
respectively [Fig.~\ref{fig:kappa}(a,b)].
However, when transforming from 3D to 2D, huge difference emerges that the
$\kappa$ of $g$-BAs (the 2D counterpart of $c$-BAs) is much lower than that of
graphene (the 2D counterpart of diamond), despite the comparable $\kappa$ of
$c$-BAs and diamond [Fig.~\ref{fig:kappa}(c)].
In the following, we perform detailed analysis to achieve fundamental
understanding on the anomalously lowered $\kappa$ of $g$-BAs compared to
$c$-BAs, diamond, and graphene.
With the uncovered underlying mechanism, the generally lower $\kappa$ of systems
in 2D than 3D form as shown in Fig.~\ref{fig:compare} can also be well
understood.

\subsection{Large scattering phase space}
%
Fig.~\ref{fig:dispersion}(a) shows the phonon dispersion of $g$-BAs, in
comparison with $c$-BAs [Fig.~\ref{fig:dispersion}(b)].
It was claimed in previous study\cite{PhysRevLett.111.025901} that the
ultra-high $\kappa$ of $c$-BAs comparable to diamond is resulted from the large
phonon band gap between acoustic and optical phonon branches together with the
bunching of the acoustic phonon branches.
The features of the phonon dispersion of $c$-BAs analyzed based on
first-principles calculations are then confirmed by experimental measurements
based on inelastic $x$-ray scattering\cite{Phys.Rev.B.2016.94.22.220303}, as
reproduced in Fig.~\ref{fig:dispersion}(b).
When transforming from 3D $c$-BAs to 2D $g$-BAs, the phonon dispersions show
some different features, which could have remarkable effect on the $\kappa$.
($i$) The phonon band gap in $g$-BAs is 13.00\,THz, which is larger than that in
$c$-BAs (9.28\,THz).
The larger phonon band gap is expected to not have a negative effect on the
$\kappa$ for the 3-phonons scattering processes considered here.
($ii$) The $z$-direction optical (ZO) phonon branch in $g$-BAs is below the
bandgap as highlighted in Fig.~\ref{fig:dispersion}(a), which could lead to more
scattering probability by coupling with acoustic phonon branches (especially
LA).
See Supplemental Note 1 and Supplemental Figure 2 for more information on the
coupling as revealed by phonon-phonon scattering channels.
Such coupling is absent in the 3D $c$-BAs.
($iii$) The bunching of acoustic phonon branches in $g$-BAs becomes weak due to
the separation of the three phonon branches.
The weakened bunching effect for acoustic phonon branches together with the
coupling with ZO phonon branch could lead to more phonon-phonon scattering, and
thus is probably responsible for the anomalously lower $\kappa$ of $g$-BAs than
$c$-BAs.


The ZO phonon branch in $g$-BAs is mainly contributed from the boron (B) atoms,
as revealed by the partial density of states ($p$DOS) in
Fig.~\ref{fig:dispersion}(c).
In fact, due to the mass difference, the optical phonon branches in both $c$-BAs
and $g$-BAs are contributed from the B atoms [Fig.~\ref{fig:dispersion}(c,d)].
With the geometry structures transformed from 3D ($c$-BAs) to 2D ($g$-BAs), the
$z$-direction vibration of B atoms is totally different due to the 2D nature of
bondings and structural symmetry, which lowers the frequency of ZO and provides
more scattering probability in $g$-BAs by strongly coupling with acoustic phonon
branches [Fig.~\ref{fig:dispersion}(a)] (Supplemental Note 1 and Supplemental
Figure 2).
Such phonomena is also observed in monolayer
GaN\cite{Nanoscale.2017.9.12.4295-4309, Phys.Rev.B.2017.95.19.195416}.

All the possible phonon-phonon scattering events quantified by the scattering
phase space are determined based on the phonon dispersions by conserving both
energy and crystal momentum with symmetry included
\cite{PhysRevB.81.085205, PhysRevB.80.125203, Phys.Rev.B.1996.53.14.9064-9073,
0953-8984-20-16-165209}
\begin{eqnarray}
\label{eq:NU}
\nonumber
\omega_j(\vec{q}) \pm \omega_{j^\prime}(\vec{q^\prime})
& = & \omega_{j^{\prime\prime}}(\vec{q^{\prime\prime}})
\ ,\\
\vec{q} \pm \vec{q^\prime}
& = & \vec{q^{\prime\prime}} + \vec{K}
\ ,
\end{eqnarray}
where $\omega$ is the frequency of phonon modes ($\hbar\omega$ is the
corresponding energy), $\vec{q}$ is the wave vector.
Normal process corresponds to $\vec{K}=0$, while Umklapp process corresponds to
$\vec{K} \neq 0$.
Fig.~\ref{fig:P3} presents the phase space of 3-phonons scattering in $g$-BAs,
in comparison with that in $c$-BAs.
The scattering phase space in $c$-BAs is small due to the large acoustic-optical
phonon band gap, which is responsible for the ultra-high $\kappa$ of $c$-BAs as
analyzed in previous study\cite{PhysRevLett.111.025901}.
However, the scattering phase space in $g$-BAs is larger than that in $c$-BAs,
especially for the frequency range of 5-10\,THz (Fig.~\ref{fig:P3}), despite the
larger gap in $g$-BAs than $c$-BAs [Fig.~\ref{fig:dispersion}(a,b)].
The enhanced scattering probability could be attributed to the weakened bunching
effect for acoustic phonon branches together with their coupling with ZO phonon
branch in $g$-BAs [Fig.~\ref{fig:dispersion}(a)], which partially explains the
anomalously lower $\kappa$ of $g$-BAs than $c$-BAs.

We also compare the phase spaces between $g$-BAs and graphene.
It is found that the overall scattering phase space in $g$-BAs is larger than
that in graphene.
Thus, the anomalously lower $\kappa$ of $g$-BAs than graphene is understandable
despite their similar structures.
However, it should be noted that the difference in the scattering phase space is
much less than one order of magnitude, which cannot fully explain the large
difference in the scattering rate (Inset of Fig.~\ref{fig:P3}) and further the
more than one order of magnitude lower $\kappa$ of $g$-BAs (137.70\,W/mK) than
graphene (3094.98\,W/mK).
Therefore, there should be some other mechanism also responsible for the
anomalously low $\kappa$ of $g$-BAs beyond the scattering phase space.

\subsection{Strong phonon anharmonicity}
%
It is well known that the phonon lifetime is governed by two factors:
phonon-phonon scattering phase space and strength.
The phonon-phonon scattering strength describes how strong is the phonon-phonon
scattering process, which is governed by the anharmonic nature of the system.
The Gr\"uneisen parameter that describes the phonon anharmonicity can be
calculated based on the change of phonon frequency with respect to the volume
change
\begin{eqnarray}
\label{eq:gamma}
\gamma = -\frac{V}{\omega} \frac{\partial\omega}{\partial V}\ .
\end{eqnarray}
Fig.~\ref{fig:anharmonicity}(a) shows the obtained Gr\"uneisen parameter of
$g$-BAs, in comparison with graphene.
The FA phonon branch possess the largest magnitude of Gr\"uneisen parameter for
both $g$-BAs and graphene, and the magnitude is larger in $g$-BAs than in
graphene, revealing the stronger phonon anharmonicity in $g$-BAs.
Thus, the more than one order of magnitude lower $\kappa$ of $g$-BAs than
graphene can be well understood by combining the larger scattering phase space
(Fig.~\ref{fig:P3}) and the stronger phonon anharmonicity in $g$-BAs.

The phonon anharmonicity can also be intuitively revealed by the potential
energy well.
To have an explicit look at the anharmonicity, the potential energy wells
(potential energy changes per atom due to the atomic displacement) of $g$-BAs
and graphene are plotted in Fig.~\ref{fig:anharmonicity}(b) for comparison.
Both the potential wells are asymmetric with respect to the positive and
negative atomic displacements, indicating the asymmetry in the ability of an
atom vibrating around its equilibrium position and the nonlinear dependence of
restoring forces on atomic displacement amplitudes, which is the direct evidence
of the anharmonicity\cite{Phys.Rev.B.2016.94.16.165445,
Angew.Chem.Int.Ed..2016.55..2-18, EnergyEnviron.Sci..2016.SnSe.Zhao}.
We further fit the calculated points for $g$-BAs and graphene, respectively,
with the polynomial
\begin{equation}
y = E_0 + a_1x + a_2x^2 + a_3x^3\ ,
\end{equation}
where $y$ is the relative energy change per atom, and $x$ is the relative
movement of atom.
The fitted parameters for $g$-BAs and graphene are listed in the table as inset
in Fig.~\ref{fig:anharmonicity}(b).
The fitted quadric ($a_2$) and cubic ($a_3$) terms correspond to the harmonicity
and anharmonicity, respectively.
The relatively smaller harmonic term in $g$-BAs reveals the weaker bonding
strength of B-As bond than C-C band in graphene, and the relatively larger
magnitude of the anharmonic term reveals stronger phonon anharmonicity in
$g$-BAs.

\subsection{Lone-pair electrons}
%
To gain deep insight into the origin of the strong phonon anharmonicity in
$g$-BAs, we further perform fundamental analysis on the electronic structures to
uncover the underlying mechanism.
We will show that the strong phonon anharmonicity in $g$-BAs is fundamentally
driven by the stereochemically active lone-pair electrons due to the special
orbital hybridization.

\begin{figure}[tb]
    \centering
    \includegraphics[width=0.98\linewidth]{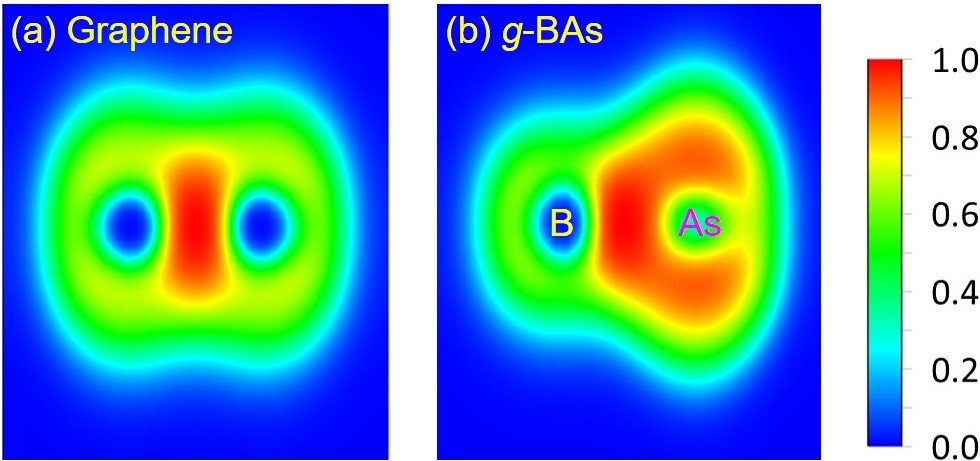}
\caption{\label{fig:ELF}
Side views of the electron localization function (ELF) for (a) graphene and (b)
$g$-BAs.
The comparison reveals the stereochemically actived lone-pair electrons in
$g$-BAs, which are non-bonding around arsenide (As) atom.
}
\end{figure}

The orbital projected electronic structures [band structure and density of
states (DOS)] of $g$-BAs are depicted in Fig.~\ref{fig:band}(a,b,c).
Direct band gap ($\sim$0.75\,eV) emerges in $g$-BAs, which is different from the
indirect band gap in $c$-BAs [Fig.~\ref{fig:band}(d,e)].
As shown in Fig.~\ref{fig:band}(a,b,c), the bonding states in $g$-BAs are
governed by the B-$s/p$ and As-$p$ orbitals.
As for B atom, all the 3 valence electrons are involved in the formation of B-As
$\sigma$ bonds due to the $sp^2$-hybridization [Fig.~\ref{fig:band}(a)].
The situation is totally different for As atom which possesses 5 valence
electrons.
The As-$s$ orbital is largely ($\sim$10\,eV) confined below the valence band,
forming an isolated band [Fig.~\ref{fig:band}(b)].
Consequently, only the As-$p$ orbitals contribute to the B-As $\sigma$ bonds.
Thus, the $s^2$ electrons in the $s^2p^3$ valence configuration of As atom do
not participate in the bonding and thus form lone-pair around the As atoms.
To have an intuitive view on the lone-pair As-$s$ electrons, we plot the
electron localization function (ELF) in Fig.~\ref{fig:ELF}.
The ELF displays the location and size of bonding and lone-pair electron, which
is powerful in interpreting chemical bonding patterns\cite{Phys.Rev.B.2016.94.16.165445}.
The ELF values range from 0 to 1, where 0 means no electron, 0.5 corresponds to
the electron-gas-like pair probability, and 1 corresponds to perfect
localization.
It is well known that in graphene the C-C $\sigma$ bonds are contributed from
the hybridized C-$s/p_x/p_y$ orbitals and the solo C-$p_z$ orbital forms the
$\pi$ bonds and the electronic Dirac cone (Supplemental Figure 3)
\cite{Balandin2012266}.
Thus, there is no lone-pair electrons formed in graphene.
By comparing the side views of the ELF between $g$-BAs [Fig.~\ref{fig:ELF}(b)]
and graphene [Fig.~\ref{fig:ELF}(a)], it can be clearly seen that there are
electrons localized around As atom that are not bonded, which are the lone-pair
electrons.

It was proposed that lone-pair electrons could lead to low
$\kappa$\cite{lone-pair1962}.
The principle underlying the concept is that the overlapping wave functions of
lone-pair electrons with valence (bonding) electrons from adjacent atoms would
induce nonlinear electrostatic forces upon thermal agitation, leading to
increased phonon anharmonicity in the lattice and thus reducing the
$\kappa$\cite{lone-pair1962,Phys.Rev.Lett..2008.101.3.035901,Phys.Rev.Lett..2011.107.23.235901,EnergyEnviron.Sci..2013.6.2.570-578,NatPhys.2015.11.12.990-991,Angew.Chem.Int.Ed..2016.55.27.7792-7796,Phys.Rev.B.2016.94.12.125203}.
Due to the orbital distribution in the same energy range and wave functions
overlap as shown in Figs.~\ref{fig:band}(a,b,c) and \ref{fig:ELF}(b), the
non-bonding lone-pair As-$s$ electrons interact with the covalently bonding
electrons of adjacent B atoms in $g$-BAs.
The interactions induce nonlinear electrostatic force among atoms when they
thermally vibrate around the equilibrium
positions\cite{2018_NE_50_425_Guangzhao_Lonepair}.
The nonlinear electrostatic force originates from the asymmetric change of the
hybridization between As-$s$ and B-$s/p$ for the atomic motion, as revealed by
the $p$DOS evolution in Supplemental Figure 4.
A more asymmetric potential energy well is induced together with the additional
nonlinear electrostatic force [Fig.~\ref{fig:anharmonicity}(b)], which leads to
the strong phonon anharmonicity in $g$-BAs [Fig.~\ref{fig:anharmonicity}(a)] and
significantly reduces the $\kappa$ of $g$-BAs [Fig.~\ref{fig:kappa}(c)].

The form of orbital hybridizations in $c$-BAs are highly consistent with those
in $g$-BAs as shown in Fig.~\ref{fig:band}, which means that lone-pair As-$s$
electrons also emerge around As atoms in $c$-BAs.
However, no strong phonon anharmonicity is induced in $c$-BAs by the lone-pair
As-$s$ electrons despite the similar orbital hybridization form as $g$-BAs,
which is due to the different bonding nature and coordination environment
between 3D and 2D geometry structures.
Due to the perfect octahedral coordination of As atoms in $c$-BAs resulted from
its cubic structure [Fig.~\ref{fig:kappa}(b)], four equivalent valence bonds are
formed.
Thus, the lone-pair electrons in $c$-BAs are stereochemically inactive, which
has no effect on the phonon anharmonicity.
In contrast, for $g$-BAs possessing planar structure, no pyramidal geometry is
formed for the B-As bonds.
Consequently, lone-pair As-$s$ electrons are located at both sides of the 2D
structure plane in $g$-BAs [Fig.~\ref{fig:ELF}(b)], which is different from that
in 3D bulk systems of $c$-BAs.
Thus, strong phonon anharmonicity exist in $g$-BAs due to the stereochemical
activity of the lone-pair As-$s$ electrons in the geometric form of planar
structure.

\subsection{Extention to other systems}

It was shown above that bond nanodesigning by changing the coordination
environment is an effective approach for realizing low $\kappa$, which would
benefit the design of thermoelectric devices with improved performance.
The approaches can also be applied to other materials beyond the BAs systems
studied here, for instance, the class of group III-V compounds (\emph{e.g.}\ BN,
AlN, GaN, \emph{etc}), where lone-pair electrons also exist
(Fig.~\ref{fig:compare}).
Strong phonon anharmonicity and low $\kappa$ could be achieved with the
stereochemically activated lone-pair electrons, which can be realized by
breaking the perfect octahedral coordination [Fig.~\ref{fig:kappa}(a,b)].
Note that the 2D structures of the systems presented in Fig.~\ref{fig:compare}
are all planar except silicene.
For the systems with buckled structures in 2D form, such as silicon \emph{vs.}\
silicene, the situation will be different due to the broken symmetry-based
selection rule for phonon-phonon
scattering\cite{PhysRevB.82.115427, Nanoscale.2017.9.12.4295-4309,
Phys.Rev.B.2017.95.19.195416}.
It is found that there exists a strong correlation between the electronegativity
difference and the $\kappa$ modulation for binary compounds
(Fig.~\ref{fig:compare}).
The effect of $\kappa$ modulation by stereochemically activating the lone-pair
electrons is weaker with a larger electronegativity difference.
The reason may lie in the contribution to phonon anharmonicity of the
electronegativity difference\cite{Nanoscale.2017.9.12.4295-4309}.
Other approaches could also have the same effects on the $\kappa$ modulation
that make the lone-pair electrons stereochemically active, such as
nanostructuring.
Alternatively, it would be also possible by substituting the atoms in ordinary
materials with special atoms that can form non-bonding lone-pair electrons, such
as nitrogen, phosphorus, arsenic, \emph{etc}.

Note that the $\kappa$ of the studied systems here does not achieve an ultralow
value, which may limit their direct applications in thermoelectrics.
However, if the approach of activating lone-pair electrons is combined with the
commonly used strategy of nanostructuring, ultralow $\kappa$ desirable for
thermoelectrics could be effectively achieved.
For example, experimental measurements have already demonstrated that the
$\kappa$ of BAs can be suppressed by the arsenic deficiency or vacancy in the
BAs sample\cite{Appl.Phys.Lett..2015.106.7.074105}
and the phonon-boundary scattering in BAs microstructures
\cite{PhysRevB.88.214303, Appl.Phys.Lett..2016.108.20.201905}.
However, the obtained $\kappa$ of BAs is still too high, which limits its
potential applications in thermoelectrics, despite its quite large Seebeck
coefficient and thermoelectric power
factor\cite{Appl.Phys.Lett..2016.108.20.201905}.
If the lone-pair electrons in the BAs system can be stereochemically activated,
the $\kappa$ could be further reduced, which would improve the thermoelectric
performance in the experimental setup.
Besides, due to the intrinsic high $\kappa$, BAs also shows promising
applications in efficient heat dissipation of electronics.
When incorporating it into conventional semiconducting devices for heat
dissipation, special attention should be paid to avoid activating the lone-pair
electrons in BAs based nanostructures for keeping the high $\kappa$.


\section{Conclusions}
%
In summary, by cutting the 3D cubic structure of $c$-BAs at the (111) cross
section, more than one order of magnitude lowered $\kappa$ is achieved in the
resultant 2D system of $g$-BAs with similar structure as graphene, which shows
that bond nanodesigning by transforming the materials into nanoscale with the
broken coordination environment could be an effective approach for realizing low
$\kappa$.
Based on the systematic study on the thermal transport properties of $g$-BAs
comparing with $c$-BAs, diamond, and graphene ($c$-BAs $\to$ $g$-BAs \emph{vs.}\
diamond $\to$ graphene), the underlying mechanism for the substantially lowered
$\kappa$ in the case of `$c$-BAs $\to$ $g$-BAs' lies in two aspects:
1)
Resulted from mass difference and 2D nature of bondings and structural symmetry,
the weakened bunching effect for acoustic phonon branches together with their
coupling with ZO phonon branch play a key role in driving large probability of
phonon-phonon scattering.
2)
Strong phonon anharmonicity is fundamentally driven by the stereochemically
actived lone-pair electrons in $g$-BAs.
Due to the special orbital hybridization, the $s^2$ electrons in the $s^2p^3$
valence configuration of As atom do not participate in the bonding but form
lone-pair instead.
When transforming from the 3D cubic structure of $c$-BAs to 2D planar structure
of $g$-BAs, the lone-pair As-$s$ electrons become stereochemically actived due
to the break of the perfect octahedral coordination of As atoms in $c$-BAs,
which leads to strong phonon anharmonicity in $g$-BAs.
Similar concept can be also extended to other systems with lone-pair electrons
beyond BAs, such as group III-V compounds (\emph{e.g.}\ BN, AlN, GaN,
\emph{etc}), where a strong correlation between $\kappa$ modulation and
electronegativity difference for binary compounds is found.
Thus, the lone-pair electrons combined with a small electronegativity difference
could be the indicator of lowering $\kappa$ through bond nanodesigning to change
the coordination environment.
The proposed approach for realizing low $\kappa$ and the underlying mechanism
uncovered in this study would shed light on future research involving novel
materials for energy applications.

\section{Computational Details}
All the first-principles calculations are performed based on the density
functional theory (DFT) as implemented in the Vienna \emph{ab initio} simulation
package (\texttt{\textsc{vasp}})\cite{PhysRevB.54.11169}.
The Perdew-Burke-Ernzerhof (PBE)\cite{PhysRevLett.77.3865} of generalized
gradient approximation (GGA) is chosen as the exchange-correlation functional
for describing boron arsenide (BAs) systems, which is produced using the
projector augmented wave (PAW) method\cite{PhysRevB.59.1758}.
Based on careful convergence test, the kinetic energy cutoff of wave function is
set as 800\,eV for all the DFT calculations.
For the 2D systems, a large vacuum spacing is necessary to hinder the
interactions arising from the employed periodic boundary conditions, which is
set as 20\,\AA\ along the \emph{out-of-plane} direction.
The Monkhorst-Pack \cite{PhysRevB.13.5188} $k$-meshes of $15\times 15\times 1$
and $2\times 2\times 1$ are used to sample the Brillouin zone (BZ) for the
structure optimizations and supercell force calculations, respectively, with the
energy convergence threshold set as $10^{-8}$\,eV.
The structure optimization is fully conducted with no limitation until the
maximal Hellmann-Feynman force acting on each atom is less than
$10^{-9}$\,eV/\AA.

For the supercell force calculations to obtain interatomic force constants
(IFCs), a $5\times 5\times 1$ supercell is constructed based on the convergence
of the phonon dispersion with respect to the supercell size.
The cutoff radius ($r^{\mathrm{cutoff}}$) introduced during the calculations of
the anharmonic IFCs is also fully tested, which is used to discard the
interactions between atoms with distance larger than a certain value for
practical purposes.
The $r^{\mathrm{cutoff}}$ of 10th nearest neighbors ($\sim$0.94\,nm) is found to
be large enough to obtain converged and reliable $\kappa$\cite{npjCM.2018.4.1.3}.
The space group symmetry properties are used to reduce the computational cost
and the translational and rotational invariance of IFCs are enforced using the
Lagrange multiplier method \cite{phonopy, PhysRevB.77.144112, PhysRevB.86.174307}.
With the anharmonic IFCs, the scattering matrix can be constructed, based on
which one can calculate all the three-phonon scattering rates and then obtain
phonon lifetime.
The Born effective charge ($Z^*$) and dielectric constant ($\epsilon$) obtained
based on the density functional perturbation theory (DFPT) are included for
taking into account of long-range electrostatic interactions.
The thickness for calculating $\kappa$ is chosen as the van der Walls diameter
(3.7\,\AA).
The $\kappa$ is obtained by solving the linearized phonon BTE using an iterative
procedure as implemented in the ShengBTE package based on the IFCs
\cite{Li20141747, PhysRevB.86.174307}.
More information can be found in Supplemental Note 2.

\section{Acknowledgments}
%
Simulations were performed with computing resources granted by RWTH Aachen
University under projects rwth0366.
M.H.\ acknowledges the start-up fund from the University of South Carolina.
Z.Q.\ is supported by the National Natural Science Foundation of China (Grant
No.\ 11847158) and the China Postdoctoral Science Foundation (2018M642776).


\section{Supplemental Information}
\noindent
\textbf{Supplemental Table} for the specific $\kappa$ of the typical systems in
3D and 2D forms;\\
\textbf{Supplemental Figures} for 1) the phonon dispersion showing quadratic
behavior of FA, 2) phonon-phonon scattering channels of FA, 3) oribitals
projected electronic structures for graphene, 4) $p$DOS evolution for $g$-BAs
due to atomic motion, and 5) normalized $\kappa$ for several typical 2D
materials;\\
\textbf{Supplemental Notes} for 1) phonon-phonon scattering channels and 2) more
information on computational methods.

\section{Data Availability}
The data that support the findings of this study are available from the
corresponding author upon reasonable request.

\bibliography{bibliography}
\bibliographystyle{elsarticle-num}

\end{document}